\documentclass[aps,prl,showpacs,preprintnumbers,amsmath,assym,
twocolumn,superscriptaddress]{revtex4}
\usepackage{amsmath}

\usepackage{graphicx}
\usepackage{epsfig}
\usepackage{bm}

\begin{document}

\begin{abstract} 
We elucidate a long-standing puzzle about the non-equilibrium
universality classes describing self-organized criticality in sandpile
models. We show that depinning transitions of linear interfaces in
random media and absorbing phase transitions (with a conserved
non-diffusive field) are two equivalent languages to describe sandpile
criticality.  This is so despite the fact that local roughening
properties can be radically different in the two pictures, as
explained here. Experimental implications of our work as well as
promising paths for future theoretical investigations are also
discussed.
\end{abstract}

\title{
Absorbing states and elastic interfaces in random media: \\ two
equivalent descriptions of self-organized criticality}

\author{Juan A. Bonachela}
\affiliation{
Instituto de F{\'\i}sica Te{\'o}rica y Computacional Carlos I,
Facultad de Ciencias, Universidad de Granada, 18071 Granada, Spain}
\author{Hugues Chat\'e}
\affiliation{CEA -- Service de Physique de l'\'Etat Condens\'e,~CEN
~Saclay,~91191~Gif-sur-Yvette,~France}
\author{Ivan Dornic}
\affiliation{
Instituto de F{\'\i}sica Te{\'o}rica y Computacional Carlos I,
Facultad de Ciencias, Universidad de Granada, 18071 Granada, Spain}
\affiliation{CEA -- Service de Physique de l'\'Etat Condens\'e,~CEN
~Saclay,~91191~Gif-sur-Yvette,~France}
\author{Miguel A. Mu\~noz}
\affiliation{
Instituto de F{\'\i}sica Te{\'o}rica y Computacional Carlos I,
Facultad de Ciencias, Universidad de Granada, 18071 Granada, Spain}

\date{\today}
\maketitle

The concept of self-organized criticality (SOC) has been proposed to
account for the emergence of scale invariance in Nature
\cite{SOC}. Its main tenet is that in the presence of slow driving and
fast dissipation, acting at infinitely-separated time scales, many
systems self-organize (without any explicit tuning of parameters) to a
critical state \cite{GG}. Some archetypical examples of SOC are
provided by sandpile toy models, in which grains are slowly added,
locally redistributed on a fast timescale whenever an instability
threshold is overcome (generating avalanches of toppling events), and
finally dissipated at the open boundaries. Upon iteration, this
process leads to a critical steady state.

SOC can be related to standard (non-equilibrium) critical phenomena by
defining the ``fixed energy ensemble'' \cite{fes-bak,FES} in which driving and
dissipation are switched off, so that the number of grains (or ``energy'') is
conserved. Using this quantity as a control parameter, a standard (i.e. non
self-organized) phase transition is observed: for large energy densities,
there is a finite density of active (toppling) sites, whereas the system ends
in a frozen (stable) state at low densities.  It has been shown that the
critical point separating these two regimes occurs at the value of the energy
density at which the system self-organizes when subjected to slow-driving and
boundary dissipation \cite{FES,Oslo-AS}. Subsequent debate has attempted to
elucidate which non-equilibrium universality class stochastic sandpiles belong
to.  As detailed below, two alternative solutions have been proposed.

Sandpiles were first related to interfaces in random media
\cite{lim-map}. In this language, the interface height, $h(x,t)$ is
the number of times a given site $x$ has toppled up to time $t$, and
frozen states correspond to pinned interfaces. The resulting
pinning-depinning transition was argued to fall in the quenched
Edwards-Wilkinson or linear interface model (LIM) class, described by
\cite{LIM-renorm,Lesch}:
\begin{equation}
\partial_t h(x,t) =  \nabla^{2}h(x,t) + F + \eta(x,h)
\label{LIM}
\end{equation}
where $F$ is a force and $\eta(x,h)$ is a quenched white noise. This
correspondence was recently proven exact between one particular sandpile model
\cite{Oslo} and one member of the LIM class \cite{Pruessner} but, in general,
it is only approximate, as some noise-correlations need to be
neglected to establish a full correspondence with Eq.(\ref{LIM}).

Alternatively, sandpile models have been rationalized as systems exhibiting an
absorbing-state phase transition \cite{FES}.  Indeed, in the fixed energy
ensemble, a stable configuration is one of the infinitely-many absorbing
states in which the system can be trapped forever, whereas activity never
ceases above the critical point. The corresponding universality class is
{\it not} the prominent directed percolation (DP) class,
but is characterized by the coupling of a DP-like activity field to a
conserved, non-diffusive, auxiliary field (the ``energy'')
\cite{FES,Romu}. Often called C-DP (or also Manna \cite{Manna}) class,
 it is characterized by the following set of Langevin equations:
\begin{equation}
\label{CDP}  
\begin{array}{rcl}
\partial_t \rho & = & a \rho - b \rho^2 + D_{\!\rho} \nabla^2 \rho
+ \omega \rho\, \phi + \sigma \sqrt{\rho}\, \eta(x,t), \\
\partial_t \phi & = & D_{\!\phi} \nabla^2 \rho \mbox{ ,}
\end{array}
\end{equation}
\noindent
where $\rho$ is the activity field, $\phi$ the background energy field, and
$\eta(x,t)$ a Gaussian white noise \cite{FES,Romu}. 

The validity of both of these alternative pictures has been
(partially) backed by numerical measurements of critical exponents
but, in general, they have not been proven to be correct so far.
But, if both of the pictures are right, a remarkable consequence
follows: depinning transition of LIM-class interfaces in random media
and the C-DP class absorbing phase transition should be equivalent,
even if they look rather different (for instance, one involves
quenched disorder and the other does not). Here, we explore this issue and
the more general question of whether 
any depinning interface universality class has an
equivalent absorbing phase transition class.

\begin{figure}
\includegraphics[width=70mm]{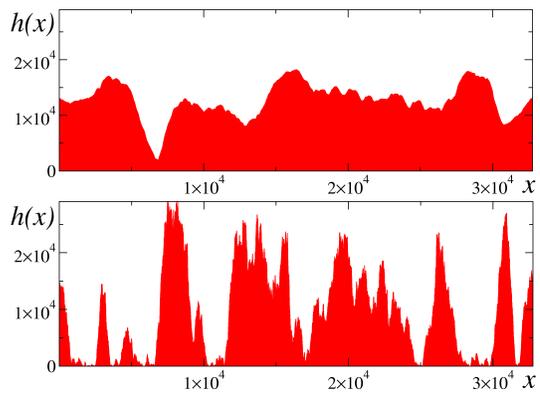}
\caption{Typical one-dimensional interfaces for a system of $2^{15}$ sites
at $t=10^5$ from flat initial conditions. Top: Leschhorn automaton
(LIM class, $\kappa\approx0.17(1)$). Bottom: interface constructed
from the Manna sandpile \protect\cite{Manna} (C-DP class,
$\kappa\approx0.43(1)$).}
\label{f1}
\end{figure}

Only a few works have approached the connections between these two
pictures. In \cite{DM}, interfaces were constructed from a DP class
model, using the cumulated local activity as the interface
height. Anomalously-rough interfaces, characterized by a positive
local-slope exponent $\kappa$ defined by $\langle (\nabla h)^2 \rangle
\sim t^{2 \kappa}$ \cite{kappa}, were found at criticality. These
anomalous interfaces are not related to any known interface class.
Focusing on SOC sandpiles, Alava and Mu\~noz \cite{MikkoMA} argued
heuristically that the LIM and C-DP classes could be identified with
each other (using also $h(x,t)=\int_0^{t}\rho(x,s)ds$ to relate
Eq.(\ref{LIM}) to Eq.(\ref{CDP})) although a one-to-one mapping could
not be rigorously established. However, this conclusion was later
challenged by Kockelkoren and Chat\'e (KC) \cite{KC} who found that
the $\kappa$ exponent takes completely different values for LIM
interfaces and for interfaces constructed from models in the C-DP
class. In particular, the constructed interfaces are anomalously-rough
below the upper critical dimension, i.e. for space dimensions $d<4$,
while LIM interfaces have $\kappa \leq 0$ (not anomalous) for
$d\ge2$. In $d=1$, both types of interfaces are anomalously rough, but
in a manifestly distinct manner, i.e. different values of $\kappa$
(Fig.~\ref{f1}).  This led KC to conclude that LIM and C-DP classes
{\it cannot} be equivalent, even if all the other recorded
``standard'' critical exponents (as $\theta$ and $z$, see definitions
below and table I, and others) take ``almost indistinguishable
values'' in these two problems, (which could, in principle, be
attributed to a numerical coincidence
\cite{KC}.)
\begin{table}
\caption{\label{t1} Some of the measured critical exponents 
of the C-DP/LIM class vs space dimension $d$
\protect{\cite{FES,Lesch,KC,Dornic,Lubeck}}.  
The A- and B-scaling values for $\kappa$ are from our own present
simulations.}
\begin{ruledtabular}
\begin{tabular}{clllll}
$d$ & $\theta$ & $z$ & $\kappa_A$ & $\kappa_B$ \\ \hline
1   &  0.13(1)    & 1.42(2) & 0.17(1)     & 0.43(1)      \\
2   &  0.51(2)    & 1.55(3) & $0^{-}$      & 0.25(2)       \\
3   &  0.77(3)    & 1.78(5) & $<0$      & 0.12(3)       \\
\end{tabular}
\end{ruledtabular}
\end{table}

While the discrepancy in values of $\kappa$ is unquestionable,
simulating directly Eqs.(\ref{CDP}) (using the method in
\cite{Dornic}) we find all the other C-DP exponents to be
indistinguishable from their counterparts in the LIM class
(Table~\ref{t1}) as well as from the corresponding values in
stochastic sandpiles.

In this Letter, we show that depinning transitions of LIM interfaces
and C-DP absorbing phase transitions are indeed two equivalent
descriptions of SOC sandpiles in spite of the discrepancies in
$\kappa$-values, which we explain.  We show, using a combination of
numerical results and scaling arguments, that there is a unique
universality class and that differences in $\kappa$-values stem from
diverging {\it local} fluctuations, inherent to the absorbing state
picture, which do not affect other long-distance properties.

Let us start by clarifying the origin of the two possible values of
$\kappa$ and the scaling laws they obey by using simple scaling
arguments. First, since $h = \int {\rm d}t \rho(t)$ its scaling
dimension is $[h] \sim t^{1-\theta}$, where $\theta$ is the density
(or interface-velocity, recalling that $\partial_t h = \rho$) critical
time-decay exponent: $\langle \rho(t) \rangle
\sim t^{-\theta}$.  This leads to $[\nabla h] \sim t^{-1/z +1 - \theta}$,
where $[\nabla^{-1}]\sim t^{1/z}$ defines the dynamical critical exponent $z$,
and therefore
\begin{equation}
\kappa_{A} = 1-\theta-\frac{1}{z} \;\;\;\;
{\rm(``A \;scaling ``\; from\; now\; on)}.
\label{A}
\end{equation}
For $d=1$, plugging the values $\theta \approx 0.13$ and $z \approx
1.42$ of the LIM or C-DP class \cite{Lubeck,FES,Lesch,KC} into this
expression leads to $\kappa \approx 0.17$ which is indeed the value
measured for LIM class interfaces \cite{Lesch}. In higher dimensions,
this scaling law yields zero (with possible logarithmic corrections in
$d=2$) or negative $\kappa$ values, i.e. no anomalous scaling, as
indeed observed in simulations \cite{KC}.

On the other hand, assuming that interface heights at adjacent sites are
asymptotically uncorrelated (this will be justified after) we have
$ \nabla h \sim \sqrt{h} \sim t^{(1-\theta)/2}$ and hence
\begin{equation} \kappa_{B} =\frac{1 - \theta}{2} \;\;\;\; {\rm(``B\;scaling'')}\;.
\label{B}
\end{equation}
KC observed that B-scaling is verified by many interfaces constructed
from microscopic models at absorbing phase transitions
\cite{KC}. Our own simulations (not shown) extend this result
to different sandpile models (simulated in the fixed energy ensemble):
the constructed interfaces of the Manna \cite{Manna}, Oslo
\cite{Oslo}, and Mohanty-Dhar \cite{MD} models show B-scaling at
criticality.

To shed some light on the physical reason for the existence of two
different $\kappa$ values, let us consider the Leschhorn automaton, a
LIM-class model showing A-scaling. It is a discretization of
Eq.(\ref{LIM}): an integer-valued height advances at each site $x$ 
following:
\begin{equation}
h(x) \rightarrow h(x) +1 \;\;{\rm iff}\;\; \nabla^2 h +F+ \eta(x,h(x)) >0,
\label{lesch}
\end{equation}
where the (discretized) Laplacian is computed using the
nearest-neighbors of $x$, and $\eta=\pm 1$ with respective
probabilities $p$ and $1-p$. Consider now the Manna sandpile, a C-DP
class model whose local rule is: if two or more grains are present at
a given site, distribute two of them randomly to the nearest-neighbors
\cite{Manna,FES}. In this case, the interface $h(x)$ encoding the number
of times a site has toppled since $t=0$ shows B-scaling, as said
before. It can be can be expressed in terms of $z_{\rm in}(x)$ and
$z_{\rm out}(x)$, the cumulated number of particles respectively
received from and given to the nearest-neighbors.  For example, in
$d=1$, $z_{\rm out}(x)= 2 h(x)$ while $z_{\rm in}$ can be expressed as
the sum of a mean flux $h(x+1) + h(x-1)$ plus a fluctuating part,
$\xi(x,h(x))$, indicating stochastic deviations from this mean
\cite{lim-map,MikkoMA,KC}. The toppling condition can be written as
$z_0(x)+z_{\rm in}(x) - z_{\rm out}(x) \geq 2$ (where $z_0(x)$ is the
initial number of grains) and thus ,expressed in terms of the
following advancement rule:
\begin{equation}
h(x) \rightarrow h(x)+1 \;\;{\rm iff}\;\; \nabla^2 h - 1 + z_0(x) +
\xi(x,h) > 0 \;.
\label{fictitious}
\end{equation}
This is very similar to Eq.(\ref{lesch}) but, as noticed in
\cite{KC}, there is a crucial difference: whereas in Eq.~(\ref{lesch})
$\eta(x,h(x))$ is a {\it bounded}, dichotomous, delta-correlated
noise, the noise term $\xi(x,h(x))$ in Eq.~(\ref{fictitious}) is a sum
of random variables (a unit is added or subtracted for each toppling)
whose amplitude, by virtue of the central limit theorem, behaves like
the square root of the average of $h(x+1)+h(x-1)$, and is therefore
{\it diverging} in time: $\langle \xi(h)^2 \rangle \sim
t^{1-\theta}$. In turn, this divergence has to be compensated by the
fluctuations of the Laplacian term in Eq.~(\ref{fictitious}) (since
the term $z_0(x)$ representing the initial condition should be
irrelevant in the long-time limit and is anyhow bounded). This is at
the origin of the strong fluctuations present in the constructed
interface (B scaling) but absent in the LIM class (see Fig.\ref{f1}).
\begin{figure}
\includegraphics[width=86mm]{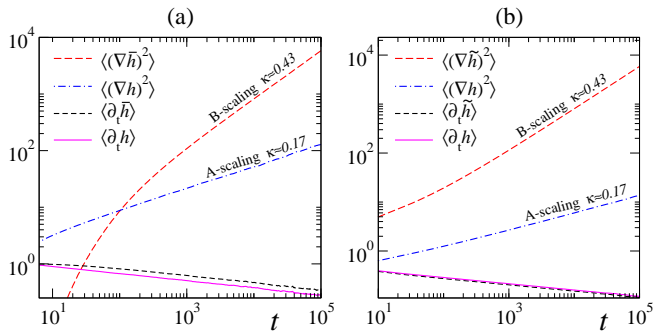}
\caption{Time series of interface squared-gradient and 
activity-density (or velocity) at criticality.
(a) Eqs.(\ref{CDP}) using
$h$, (A-scaling, $\kappa \approx 0.17(1)$) or $\bar{h}$ (B-scaling,
$\kappa \approx 0.43(1)$ ). Parameters: $L=2^{15}$,
$b=w=\sigma^2/2=1$, $D=D_{E}=0.25$, time-mesh $0.1$; critical point
$a_{c}=0.86452(5)$.  (b) Leschhorn automaton with $L=2^{15}$,
$F=0$, at the critical point $p=0.80078(5)$, and using $h$ as in
Eq.(\ref{LIM}) (A-scaling, $\kappa \approx 0.17(1)$) or $\tilde{h}$
as in Eq.(\ref{mod-LIM}) (B-scaling, $\kappa \approx 0.43(1)$) with
$p_c=0.76935(5)$.}
\label{f2}
\end{figure}
At this point, one clearly appreciates the qualitative difference between LIM
and C-DP-constructed interfaces, which occurs despite of the fact that both
classes share numerically-indistinguishable (standard) critical exponents, as
said before. We now show that this is not the end of the story, and
that we can construct A-scaling interfaces from C-DP class models, as well as
modify LIM-class models to obtain B-scaling.

Our first evidence showing that A-scaling and B-scaling can both be
compatible with a unique universality class was provided by numerical
integrations of Eqs.(\ref{CDP}). Constructing an interface, as before,
via $h(x,t) = \int{\rm d}s \rho(x,s)$ where $\rho$ is the continuous
activity field, we obtain clear A-scaling (Fig.~\ref{f2}), in contrast
with the B-scaling heretofore always observed with microscopic
models. Next, mimicking
microscopic models in which the interface advances by one {\it unit}
whenever a site is active, we constructed a different interface for
Eqs.(\ref{CDP}) through $\bar{h}(x,t) = \int_0^t {\rm
d}s\Theta(\rho(x,s))$ where $\Theta$ is the Heaviside step function;
this new interface advances by one unit whenever there is some
non-zero activity, regardless of its magnitude.  Strikingly, B-scaling
is then observed (Fig.~\ref{f2}). Thus both A- and B-scaling
interfaces have been constructed at the {\it same} absorbing phase
transition point: they correspond to slightly different {\it
observables}. We have reached a similar conclusion for the Oslo
sandpile model by taking advantage of a recent result by Pruessner
\cite{Pruessner} who constructed an exact mapping between this particular
sandpile and the Leschhorn automaton.  The local rule for $d=1$ is as
follows: distribute one grain to each nearest-neighbor whenever the
local height threshold is passed, this local threshold being randomly
reset to be $1$ or $2$ grains after each toppling. To achieve an exact
mapping, Pruessner showed that it is crucial to use the more
symmetrical $h^\dag(x)=h(x+1)+h(x-1)$, i.e.  the accumulated number of
times a given site has been charged by its neighbors rather than the
accumulated activity at the site itself (the definition of
$h$). Indeed, using $h^\dag(x)$ for the Oslo model eliminates the
diverging noise in Eq.(\ref{fictitious}) and yields A-scaling in
numerical simulations, whereas using $h(x)$ we observe B-scaling
(results not shown). We have been able to extend easily this procedure
to other sandpiles as the Mohanty-Dhar one \cite{MD}. For other
sandpiles with less symmetric redistribution rules as, for instance,
the Manna one \cite{Manna} this can be much more complicated. In this
last, the $2$ toppling grains at any site can go to the same
neighbor. This introduces an extra noise that needs to be subtracted
by an appropriate (and intricate) definition of the height variable
(different from $h$ and $h^\dag$) to get rid of intrinsic local
fluctuations and disentangle the hidden A-scaling.  These results show
that a unique universality class is compatible with different values
of $\kappa$, depending on microscopic details
and/or the definition of the height variable; {\it different $\kappa$
exponents correspond to different, though very similar,
observables}. Note also that the same definition of the interface can
lead to the two different types of scaling depending on the rules of
sandpile under study.

To close the loop, we now show that for standard interface
depinning transitions it is also possible to generate two different
$\kappa$ values without affecting other exponents. Let us define
$\partial_{t}\tilde{h}(x,t)=\partial_{t}h(x,t) (1+\sigma(x,{h})),$
where the quenched noise $\sigma$ is $0$ or $1$ with probabilities $p$
and $1-p$, respectively, so every time the original interface
advances, a noise variable is added to $\tilde{h}$. Both interfaces
are related by $\tilde{h}(x,t)=h(x,t)+\tilde{\sigma}(x,\tilde{h})$
where $\tilde{\sigma}(x,\tilde{h})$ is an accumulated noise summing up
all values of $\sigma(x,h)$ at $x$ up to height $h$. In terms of
$\tilde{h}(x,t)$, Eq.(\ref{LIM}) becomes
\begin{equation}
\partial_{t}\tilde{h}=[\nabla^{2}(\tilde{h}(x,t)-
\tilde{\sigma}(x,\tilde{h}))+\eta(x,\tilde{h})]
\times [1+{\sigma}(x,\tilde{h})].
\label{mod-LIM}
\end{equation}
Simulating Eq.(\ref{mod-LIM})
we observe B-scaling for the $\tilde{h}$-interface, while removing the
$\tilde{\sigma}$-noise we readily recover A-scaling for the
$h$-interface (Fig.~\ref{f2}), while all the other exponents coincide
for both interfaces. Na{\"i}ve power-counting for Eq.(\ref{mod-LIM})
shows that $\nabla^2 \tilde{\sigma}(x,\tilde{h})$ is an irrelevant
higher-order noise, and so is the term $\sigma(x,\tilde{h})$ added to
$1$ \cite{LIM-renorm}. Hence, upon coarse-graining, Eq.(\ref{mod-LIM})
flows towards the standard LIM renormalization group fixed point
\cite{LIM-renorm}, justifying that all {\it universal} critical
exponents should coincide in Eq.(\ref{LIM}) and Eq.(\ref{mod-LIM}) in
accordance with our numerical results. The inclusion of the extra
noise (a higher-order correction to scaling) is thus able to alter the
value of $\kappa$, intensifying anomalous behavior, but not standard
long-distance critical exponents, which are controlled by a unique
renormalization group fixed point.  The two different values of
$\kappa$ correspond to two different height variables, $h$ and
$\tilde{h}$, differing by a diverging noise. A full understanding,
within the renormalization group perspective, of how local anomalous
roughening properties are affected by an otherwise irrelevant noise
remains a challenging task.

The fact that a given universality class can be compatible with
different types of local roughening (amenable to experimental
analysis) is also of interest in general studies of fluctuating
interfaces \cite{kappa}. Indeed, one can show that $\chi$ and
$\chi_{loc}$ being respectively the global and local saturation
roughness exponents are related by $\chi = \chi_{loc} + z \kappa$
\cite{kappa}, which we have numerically verified for all models
studied here. Given that our results indicate that local roughening
properties, as encoded by $\kappa$ (or $\chi_{loc}$), can adopt
different values, depending on the presence or absence of local
fluctuations, while $\chi$, as all other non-local exponents, is
universal, it is intriguing that experiments on the propagation of
fracture cracks in wood \cite{fracture_exp} seem to lead to the
opposite conclusion. See also the nice recent experiments measuring
the exponents reported here for self-organized superconductors
\cite{Wijn}.  More experimental
work along these lines would be most welcome, since experimental
realizations of absorbing phase transitions (even for the paradigmatic
DP class) are barely existing
\cite{Haye_DPexp}.

On the theoretical side, since the C-DP class and the LIM class are
two faces of the same problem, both share the same upper critical
dimension $d_{\rm u}=4$. This result is in contradiction with the
perturbative approach for Eq.(\ref{CDP}) in \cite{FvW}. In fact,
field-theoretical treatments of the LIM class demand a functional
renormalization group calculation
\cite{LIM-renorm}, and one finds that the correlator of the quenched
noise develops a cusp in the $h$-variable. Given our results, it would
be very interesting to sort out the analogue of all this in the
absorbing phase transition picture, as well as attempting a
non-perturbative renormalization group approach for such a case. 

In summary, depinning transitions of linear interfaces in random media
and absorbing phase transitions in the C-DP class are two equivalent
descriptions of sandpile self-organized critical points.  This
clarifies the issue of universality in stochastic sandpiles and the
connection of SOC to standard non-equilibrium phase transitions and
opens the door to new and exciting research lines.


\end{document}